\documentclass[twocolumn,prl,aps,longbibliography]{revtex4-2}
\usepackage{amsmath}
\usepackage{amssymb}
\usepackage{todonotes}
\usepackage[unicode=true,colorlinks=true,citecolor=blue,urlcolor=blue]{hyperref}
\usepackage{bm}
\usepackage{setspace}
\renewcommand {\phi}{{\varphi}}
\newcommand {\rmi}{{\rm i}}
\newcommand {\rmd}{{\rm d}}

\newcommand {\e}{{\rm e}}
\newcommand {\eps}{\varepsilon}

\newcommand{\gf}{\gamma_{\to}}
\newcommand{\gb}{\gamma_{\leftarrow}}
\renewcommand{\Re}{\mathop\mathrm{Re}\nolimits}
\renewcommand{\Im}{\mathop\mathrm{Im}\nolimits}

\let\oldaddcontentsline\addcontentsline
\renewcommand{\addcontentsline}[3]{}%

\begin{document}

\title{Chiral Dissociation of Bound Photon Pairs for a Non-Hermitian Skin Effect}

\author{Jiaming Shi}
\email{jiaming.shi@weizmann.ac.il}
\author{Alexander N. Poddubny}
\email{poddubny@weizmann.ac.il}
\affiliation{Department of Physics of Complex Systems, Weizmann Institute of Science, Rehovot 7610001, Israel}

\begin{abstract}
We theoretically study the bound states of interacting photons propagating in a waveguide chirally coupled to an array of atoms. We demonstrate that the bound photon pairs can concentrate at the edge of the array and link this to the non-Hermitian skin effect. Unlike  tight-binding non-Hermitian  setups, the bound states in the waveguide-coupled array exhibit infinite radiative lifetimes when  the array has an infinite size. However, in a finite array, non-Hermiticity and localization of bound pairs emerge  due to their chiral dissociation into scattering states. Counterintuitively, when the photons are preferentially emitted to the right, the bound pairs are localized at the left edge of the array and vice versa.
\end{abstract}

\date{\today}
\maketitle

\textit{Introduction. }
Non-Hermitian skin effect (NHSE) has now become the paradigmatic  example of the topologically nontrivial effects of loss or gain in optical and condensed matter systems~\cite{Yao2018,Lee2016,Torres2018,Kunst2018,Zhang_2020,Slager2020,Okuma2020,Bergholtz2021}. Namely, if the energy spectrum $\varepsilon(k)$ as a function of the momentum $k$ forms a loop in a complex plane, the macroscopic fraction of the eigenmodes in the finite-size structure with  open boundary conditions is localized at the edge. Typically, such complex-valued loops of the energy spectrum $\varepsilon(k)$  arise due to local loss or gain, or unequal forward and backward  tunneling links \cite{Hatano1997}. The loss can also be nonlocal; for example, it can result from the collective spontaneous emission of photons into the far field~\cite{Poddubny2024skin}. Recently, NHSE has also been investigated in the quantum setup in the presence of interactions \cite{Lee2021,Shen2022,Poddubny2023skin,Brighi_2024,wang2024exoticlocalizationbodybound}. In particular, it has been shown that two interacting particles with different masses can get localized  at an edge, even if the structure is Hermitian and time-reversal symmetry is preserved~\cite{Poddubny2023skin}. These results suggest the presence of other unchartered mechanisms for NHSE in interacting systems.

\begin{figure}[b]
\centering
\includegraphics[width=\linewidth]{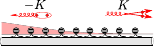}
\caption{Schematics of an edge bound state of photon pairs in an array of atoms chirally coupled to a waveguide. The bound pair experiences dissociation when propagating to the right, leading to its localization at the left edge. }
\label{fig:schematics}
\end{figure}

In this Letter, we consider the non-Hermitian skin effect for a composite particle.  We  focus on two photons  bound via atom-mediated interactions, 
as illustrated in Fig.~\ref{fig:schematics}. Such bound pairs  can be realized for cold atom systems~\cite{Winkler2006,Firstenberg2013,Prasad2020} and have also been predicted for  microwave photons coupled to  superconducting qubits setup~\cite{Zhang2019arXiv}.   Contrary to the previously studied mechanisms, here, the non-Hermiticity, essential for NHSE, is provided neither by loss nor gain but just by an intrinsic feature of the bound pair: its dissociation into  scattering states.

\begin{figure}[t]
\centering
 \includegraphics[width=\linewidth] {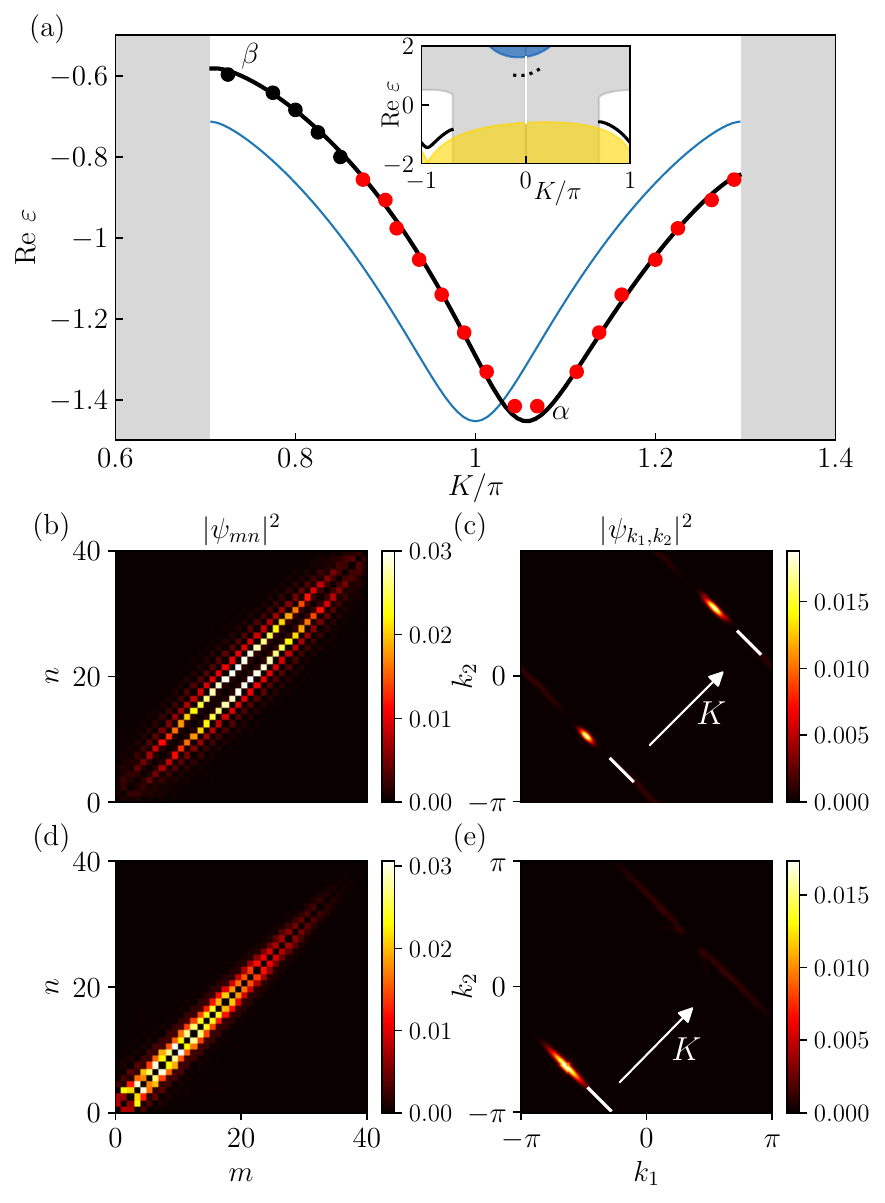}
\caption{(a) Bound pair dispersion law $\varepsilon_{\rm pair}(K)$ in the chiral (thick black curve) and non-chiral (thin blue curve) structure. The shaded gray area illustrates the scattering states continuum, where bound states do not exist. Circles show the energies and the corresponding {center-of-mass} momenta for eigenstates in the finite-size structure; black circles correspond to unidirectional states. The inset shows the same chiral dispersion curve in a larger range; blue, gray,  and yellow areas show continuums with two upper-branch polaritons,  one upper-branch and one lower-branch, and two lower-branch, respectively.
(b-e) Real-space (b,d) and reciprocal space (c,d) distributions for the two states $\alpha$ (c,d) and $\beta$ (d,e) indicated in (a).  Arrows in (c,e) illustrate the center of mass propagation direction $K$, and white lines indicate the values of $K=k_1+k_2$ extracted from   $|\psi_{k_1,k_2}|^2$\:.
The calculation  has been performed for $\varphi=0.35\pi$, $N=40$, $\xi=0.7$ and $\gamma_{\rm 1D}=1$. 
}
\label{fig:bound-dispersion}
\end{figure}

Before proceeding to the rigorous theoretical model,  we  present a qualitative explanation of  the proposed mechanism. We start with 
the dispersion relation for the bound pair,  $\varepsilon_{\rm pair}(K)$, schematically illustrated in Fig.~\ref{fig:bound-dispersion}(a), here $K={k_1+k_2}$ is the center-of-mass momentum. The thick black and thin blue curves correspond to chiral and non-chiral structures, respectively. The black curve lacks mirror symmetry; $\varepsilon_{\rm pair}(\pi+K)\ne \varepsilon_{\rm pair}(\pi-K)$.  Chirality is an essential but not sufficient ingredient for this NHSE mechanism. The second essential ingredient is the presence of a continuum of scattering states, represented by the shaded area in Fig.~\ref{fig:bound-dispersion}(a). This means that the bound states dissociate and cease to exist when their momentum $K$ is in a certain range. Our main finding is a range of the energies belonging to the bound state spectrum $\varepsilon_1\ldots \varepsilon_2$ in Fig.~\ref{fig:bound-dispersion}(a), indicated by the black circles, where $K-\pi$ takes only negative values and group velocity has only a negative sign. The ``unidirectional'' range is a consequence of {both the chirality that makes  $\varepsilon_{\rm pair}(\pi+K)\ne \varepsilon_{\rm pair}(\pi-K)$ and the scattering continuum} that renders the dispersion law $\varepsilon_{\rm pair}(K)$ non-analytical. Without such a continuum, there has to be an even number  of the solutions $K$ for the equation 
$\varepsilon_{\rm pair}(K)=\varepsilon_0$ because of the bound state branch  continuity and periodicity, $\varepsilon_{\rm pair}(K)=\varepsilon_{\rm pair}(K+2\pi)$. One can then expect that the bound eigenstates in the finite structure
will become standing waves of the type
\begin{equation}\label{eq:standing wave}
\psi_n=A\e^{\rmi K_1n}+B\e^{\rmi K_2n}\:,  
\end{equation}
where $K_{1,2}$ is the pair of solutions and $n$ is the center of mass coordinate.
The scattering continuum can destroy one of these two solutions, making the bound state unidirectional. We will demonstrate that the  bound states, corresponding to the unidirectional spectral range, will be localized at the edge under open boundary conditions, which can be interpreted as an analog of NHSE, as illustrated in Fig.~\ref{fig:schematics}. This loss mechanism is rather special: in an infinite structure, the portions of the dispersion curve $\varepsilon_{\rm pair}(K)$ corresponding to bound states outside the scattering continuum are real-valued, meaning the bound pairs have an infinite lifetime for a fixed value of $K$. This is a qualitative difference  of the considered {nonlocal} model from non-reciprocal tight-binding Bose-Hubbard-type models~\cite{Lee2021,Shen2022,Brighi_2024,wang2024exoticlocalizationbodybound}, where $\varepsilon_{\rm pair}(K)$  is complex-valued { for every $K$}. However, in a finite array, considered bound states acquire a finite lifetime and become localized at the structure's edge.

{\it Waveguide quantum electrodynamics model.} We now discuss the specific implementation of the proposed dissociation-driven NHSE in an array of two-level atoms chirally coupled to a waveguide, see Fig.~\ref{fig:schematics}. The single-particle eigenstates are polaritons, formed by the hybridization of photons with a linear dispersion relation $\omega_k = c|k|$ and atoms with resonance frequency $\omega_0$. Due to chirality, the Rabi splittings at the two avoided crossings of the polariton dispersion curve for positive and negative $k$ have different values (see Fig.~S1 in \footnote{Online Supplementary Materials}). 

We focus on the bound double-excited states $|\psi\rangle=\sum_{m,n}\psi_{mn}\sigma_m^\dag\sigma_n^\dag |0\rangle$ where the  wavefunction amplitude $|\psi_{mn}|$ decays with the distance between the two excitations $|m-n|$. Here, $\sigma_{m}^\dag$ are the raising operators for the atoms. The
structure of double-excited states  in this model, known for a long time~\cite{Yudson1984,Shen2007,Yudson2008}, has received significant attention in the  last several years~\cite{Zhang2019arXiv,Mahmoodian2020,Zhong2020,alex2020quantum,Poddubny2019quasiflat,Poshakinskiy2021chaos,Fedorovich2022,bakkensen2021}, including the study of the chiral systems~\cite{Kornovan_2021,bakkensen2021,Calajo2022,Iversen_2022}, see in particular Refs.~\cite{bakkensen2021,Calajo2022} for the chiral bound states, that were also indirectly observed in experiment~\cite{Prasad2020}. However, the general connection of the chiral bound pair dissociation to the non-Hermitian skin effect manifested as a spectral range of edge-localized bound states, to the best of our knowledge, has remained completely unexplored.

The effective non-Hermitian Hamiltonian of the structure is given by
$H=\sum_{m,n=1}^NH_{m,n}\sigma_m^\dag\sigma_n^{\vphantom{\dag}} $~\cite{Sheremet2023,Poddubny2024skin}
\begin{equation}\label{eq:H}
H_{m,n}=\omega_0\delta_{m,n}-\rmi\begin{cases}
 \gf \e^{\rmi\varphi |m-n|}, &m>n\\
 \frac{\gf+\gb}{2}, &m=n\\
 \gb \e^{\rmi\varphi |m-n|}, &m<n
\end{cases}\:.
\end{equation}
Here, $\omega_0$ is the atom resonant frequency, $ \gf=2\gamma_{\rm 1D}/(1+\xi)$  and  $ \gb=2\gamma_{\rm 1D}\xi/(1+\xi)$, 
are the spontaneous decay rates into the waveguide in the forward and backward directions, respectively. 
The parameter $\varphi=\omega_0d/c$ represents the phase gained by light propagating the distance $d$ between adjacent atoms with the velocity $c$. The Hamiltonian
Eq.~\eqref{eq:H} is written in the Markovian approximation with traced-out photon modes, which is  valid for $\gamma_{\rm 1D}\ll \omega$.
The chirality of the structure is quantified by the parameter $\xi$, that is the ratio of emission rates in the forward and backward 
directions. The non-chiral case corresponds to $\xi=1$. The array features long-ranged waveguide-mediated coupling.

When the structure has translational symmetry, the center-of-mass wave vector $K$ is a good quantum number and the two-photon wave function $|\psi\rangle = \sum_{m,n} \psi_{mn}\sigma_m^\dag \sigma_n^\dag|0\rangle $ can be sought in the form
\begin{equation}
    \psi_{mn} = \e^{\rmi K (m+n)/2} \chi_{m-n}\:.
\end{equation}
The bound pair dispersion law $\varepsilon_{\rm pair}(K)$  can be found by diagonalizing the following Hamiltonian~\cite{Calajo2022}:
\begin{align}\label{eq:4}
        \hat{H}_K = -\rmi \sum_{r,r'>0} \sum_{\eta=\pm1} &\Bigl[\gb \e^{\rmi(\varphi+\frac{K}{2}) |r +\eta r'|} \\&+ \gf \e^{\rmi(\varphi-\frac{K}{2}) |r + \eta r'|} \Bigr]\sigma_r^\dagger \sigma_{r'}^{\vphantom{\dagger}}\:,\nonumber
\end{align}
and $\hat{H}_K \chi = 2\varepsilon_{\rm pair}(K) \chi$\:.
 
The results of the calculations  are shown by the curves in Fig.~\ref{fig:bound-dispersion}(a). 
We focus on a single branch of the bound states that, in the limit of vanishing chirality, $\xi\to 1$ has the energy $\eps_{\rm pair}=2\gamma_{\rm 1D}\cot 2\varphi$~\cite{Poddubny2019quasiflat} at $K=\pi$. This branch is formed by binding two polaritons from the  different dispersion branches, the upper one with $|k|<\varphi$ and the bottom one with $|k|>\varphi$. Due to the avoided crossing at $k=\varphi$, where the single-polariton dispersion diverges in the Markovian approximation~\cite{Sheremet2023, Note1}, the bound states exist only for ${|K|>2\varphi}$.  For $|K|<2\phi$, they dissociate into the continuum (gray areas in the left and right parts of Fig.~\ref{fig:bound-dispersion}a) and become resonance states~\cite{bakkensen2021}.
The  inset of Fig.~\ref{fig:bound-dispersion}(a) also presents the  two-photon dispersion in a broad energy and wave vector range.

Near the  edge of the Brillouin zone, the pair dispersion law can be approximated by a Taylor series:
\begin{equation}\label{eq:epair}
    \varepsilon_{\rm pair}(K)\approx
    \varepsilon_{\rm pair}(\pi)+\alpha(K-\pi)+\frac{(K-\pi)^2}{2m}\:.
\end{equation}
Here, $m$ is the effective mass of the bound pair. It has been calculated analytically for $\xi=1$ in Ref.~\cite{Poddubny2019quasiflat}. The linear-in-$K$ term appears only for the nonvanishing chirality. In the lowest order in $\xi-1$ this term can be calculated by the usual $\bm K\cdot \bm p$ perturbation theory ~\cite{Note1}:
\begin{equation}\label{eq:alpha}
    \alpha(\xi)\approx \frac{1-\xi}{8}\frac{\cos 3\varphi}{\cos^5\varphi}\:.
\end{equation}
As a result of the chirality, the extremum of the dispersion curve shifts from the point $K=\pi$ and the curve loses mirror  symmetry  {around the point $K=\pi$; $\varepsilon_{\rm pair}(\pi+K)\ne \varepsilon_{\rm pair}(\pi-K)$}; compare black and blue curves in Fig.~\ref{fig:bound-dispersion}(a). As discussed in the introduction, the breakdown of mirror symmetry and the absence of the bound states for $K<2|\varphi|$ means that there is a unidirectional part of the dispersion curve when the equation $\eps=\varepsilon_{\rm pair}(K)$ has only one solution. We stress that the pair dispersion in the infinite periodic structure stays real outside the gray area. This distinguishes considered waveguide setup from  systems with  local losses or gain at each site, such as  the Hatano-Nelson model~\cite{Hatano1997},
that have complex spectrum under the periodic boundary conditions.

We now  calculate the eigenstates $\psi_{mn}$ in the finite-size array with $N$ atoms. This is done by direct numerical diagonalization of the Hamiltonian $H$ Eq.~\eqref{eq:H}, without assuming translational symmetry. For each two-photon amplitude $\psi_{mn}$ we also calculate the corresponding Fourier transform $\psi_{k_1,k_2}=\sum_{m,n}\psi_{mn}\exp[-\rmi (k_1m+k_2n)]$.
Panels {(b-e)} in Fig.~\ref{fig:bound-dispersion} present the two-photon amplitudes (b,d) and their Fourier transforms (c,e) for two characteristic bound states $\alpha$ and $\beta$, indicated in the dispersion branch in Fig.~\ref{fig:bound-dispersion}(a). For each state, we  numerically extract the corresponding center of mass wave vectors 
$K={k_1+k_2}$  from the {positions of the maxima} of the Fourier transform $|\psi_{k_1,k_2}|^2$. The corresponding momenta are indicated in Fig.~\ref{fig:bound-dispersion}(c,e) by thin white lines perpendicular to the $K$ axis, shown by white arrows. Next, we put a circle at the energy equal to $\Re \eps$ and the extracted momentum  $K$ on the dispersion plot Fig.~\ref{fig:bound-dispersion}(a). The results demonstrate a good agreement between the dispersion of the pairs in the infinite structure with the periodic boundary conditions and the dispersion of the same pairs in a finite structure.
This agreement can be interpreted as the pair being quantized in the finite array as a single composite particle, given that the array size  $N$  significantly exceeds the pair size. This behavior is analogous to exciton quantization as a whole in wide semiconductor quantum wells~\cite{IvchenkoPikus}.

For most of the bound states, there exist two maxima $K_1$ and $K_2$ in the Fourier transform (red circles in Fig.~\ref{fig:bound-dispersion}a). This is quite natural:  the bound pair forms a standing wave in the finite-size array, and a standing wave has two Fourier components.  We term such states ``bidirectional''. Our main finding is the presence of  ``unidirectional'' bound states with just one distinct maxima for $K$. These are indicated  by {black circles} in Fig.~\ref{fig:bound-dispersion}(a) and correspond to the unidirectional part of the bulk dispersion branch. We  show in Fig.~\ref{fig:bound-dispersion}(d) that these ``unidirectional'' states are concentrated at the edge of the structure, in stark contrast to the bidirectional states.
 In the  non-chiral case the unidirectional states are absent, see supplementary Fig.~S3 in~\cite{Note1}.
{Counterintuitively, the bound states in the chiral structure are concentrated at the left edge, even though the calculation has been performed for $\gamma_\to>\gamma_{\leftarrow}$. 
 Opposite edge localization can be explained by the negative sign of the polariton group velocity. It  may still seem contradictory since for  $\xi \ll 1$ ($\gamma_\to \gg \gamma_{\leftarrow}$), all the states should be localized on the right edge~\cite{Poddubny2024skin}. We  show in \cite{Note1} that the binding energy tends to zero in this limit, and the considered bound state disappears by merging with the continuum.  This resolves the apparent contradiction.}

{\it Non-Hermitian skin effect for bound states.} The concentration of the bound pairs on the structure edge is a strong indication of the non-Hermitian skin effect. The specific mechanism of non-Hermiticity and losses requires a more analytical investigation. To this end, we now present an effective model for the composite bound states that allows us to distill the mechanisms for the results in Fig.~\ref{fig:bound-dispersion}.
We consider  an effective pair Hamiltonian $\mathcal H$ given by
\begin{equation}\label{eq:Hcomp}
    \mathcal  H=\sum\limits_{n} t(\e^{\rmi \Phi} a_n^\dag  a_{n+1}^{\vphantom{\dag}}  +
    \e^{-\rmi \Phi } a_{n+1}^\dag  a_{n}^{\vphantom{\dag}}) -\rmi \sum\limits_{n,n'}U_{nn'}a_n^\dag a_{n'}^{\vphantom{\dag}}\:,
\end{equation}
where $a_n$ is an annihilation operator for a single bound pair. 
The  first term in Eq.~\eqref{eq:Hcomp}  describes the chiral dispersion of a bound pair moving on a one-dimensional lattice. The parameter  $t$ is the real tunneling amplitude, and the phase $\Phi$ characterizes the chirality of the tunneling. This phase introduces an asymmetry in the hopping between neighboring lattice sites, breaking  the time-reversal symmetry.
{The chirality itself however is not sufficient, since the first term in Eq.~\eqref{eq:Hcomp} with complex-conjugated forward- and backward- hopping amplitudes is  still  Hermitian.}
\begin{figure}[t!]
\centering
\includegraphics[width=\linewidth]{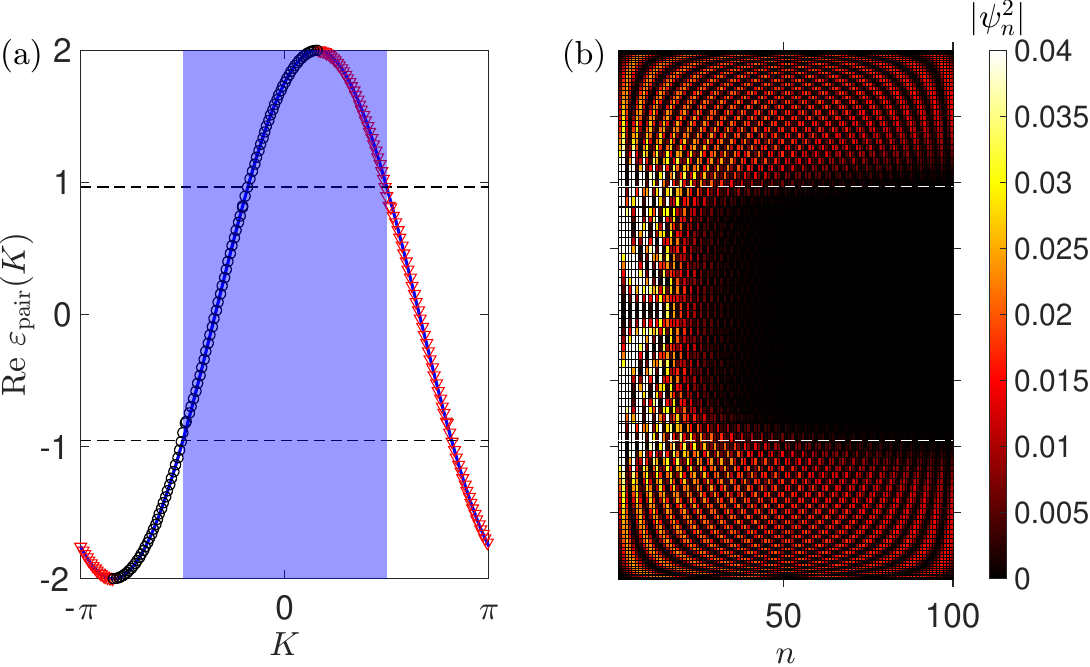}
\caption{(a) Solid line shows the dispersion law $\Re\varepsilon_{\rm pair}(K)$ calculated  following Eq.~\eqref{eq:Eloop}.  Blue shaded area illustrates the range of momenta $-2\varphi<K<2\varphi$ with a large loss.
Triangles and circles show the momenta corresponding to the eigenstates in the finite structure with the open boundary conditions.
(b) Spatial probability distribution for all the eigenstates in the finite structure.
Horizontal dashed lines in (a,b) indicate the unidirectional range of energies corresponding to the solutions concentrated at the edge.
Calculation has been performed for the following values of parameters: $t=1$,
$\Phi = -0.5$, 
$2\phi = \pi/2$,
$\Gamma = 0.3$, 
$\sigma = 0.05$, and $N=100$. 
}
\label{fig:1D}
\end{figure}
We also introduce the second, nonlocal and non-Hermitian, term in Eq.~\eqref{eq:Hcomp} $\propto U_{nn'}$, that is more complicated. This term phenomenologically describes the loss the bound pair experiences when its center-of-mass momentum $K$ is in a certain range $-2\varphi<K<2\varphi$. The  $K$-dependent loss $U(K)$ can be formally implemented by a Fourier transform 
\begin{equation}
    U_{nn'}=\int\limits_{-\pi}^{\pi}\frac{\rmd K}{2\pi} \e^{\rmi K(n-n')}U(K)\:.
\end{equation}
 We assign $U(K)$ to be equal to $\Gamma$ for $|K|<2\varphi$ and 0 otherwise. In our actual numerical calculation, instead of the step-functions, we use the sigmoidal-type functions, that is 
\begin{equation}
    U(K)=\frac{\Gamma}{\pi} \left(\arctan\frac{K+2\varphi}{\sigma}-\arctan\frac{K-2\varphi}{\sigma}\right)\:.
\end{equation}
Here, the parameter $\sigma$ is the finite step width for the  loss function in the momentum space.

The results of calculating the eigenstates of the effective pair Hamiltonian Eq.~\eqref{eq:Hcomp} are shown in  Fig.~\ref{fig:1D}. Panel (a) presents the  real parts of the energies of the eigenstates. The thin curve shows the pair spectrum in the infinite structure with the periodic  boundary conditions, described by the following equation
\begin{equation}\label{eq:Eloop}
    \varepsilon_{\rm pair}(K)=2t\cos(K+\Phi)-\rmi U(K).
\end{equation}
The  dispersion curve manifests the chirality of the structure: it has the reflection symmetry around the point $K=-\Phi$, shifted from the origin.

The triangles and circles present the results of the numerical calculation for a finite structure. Inspired by the results in Fig.~\ref{fig:bound-dispersion} we expect that each eigenstate $\sum_{n=1}^N\psi_na_n^\dag |0\rangle$  of Eq.~\eqref{eq:Hcomp} can be approximated by a sum of two plane waves, see Eq.~\eqref{eq:standing wave}.
 In order to extract the real parts of the wave vectors $K_{1,2}$ we  calculate the Fourier transform of the wavefunction amplitude 
$\psi(K)=\sum_n\psi_n\exp(-\rmi Kn)$ and assign $K_{1,2}$ to the  two maxima of $|\psi(K)|^2$. Next, we show the real parts of the eigenenergies $\varepsilon_{\rm pair}$ versus $K_{1}$ ($K_2$) by black triangles (red cicles). The results perfectly agree with the spectrum in the infinite structure~Eq.~\eqref{eq:Eloop} (blue curve).
\begin{figure}[tb!]
\centering
\includegraphics[width=\linewidth]{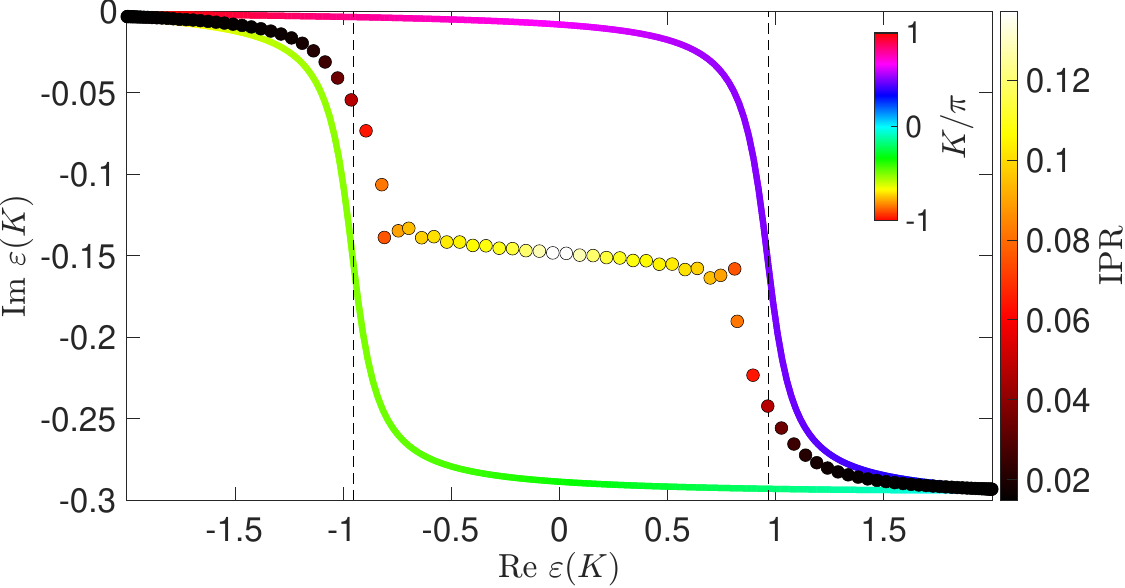}
\caption{Colored curve shows the complex pair energy dispersion $\varepsilon_{\rm pair}(K)$ calculated following Eq.~\eqref{eq:Eloop}.  The color encodes the value of $K\in [-\pi\ldots \pi]$. Filled circles present the energies of the eigenstates for the finite structure. The circle's color encodes the eigenstate's inverse participation ratio (IPR) and the curve color encodes Re~$K$. Calculation parameters are the same as in Fig.~\ref{fig:1D}.
}
\label{fig:loop}
\end{figure}

Similarly to Fig.~\ref{fig:bound-dispersion}, 
we use blue shading in Fig.~\ref{fig:1D}(a) to indicate the region $|K|<2\varphi$ of large loss.
The spectral region between the two horizontal lines is the unidirectional range of energies $-0.8\le\varepsilon\le 1.1$, where the imaginary parts for the two solutions differ significantly, and we expect the modes to accumulate at the structure edge. As expected, the unidirectional energy range indeed corresponds to the eigenstates concentrated on the edge. We have confirmed that the localization is exponential and can be switched from the left edge to the right one by changing the sign of the chirality parameter $\Phi$~(Fig.~S4 in \cite{Note1}). 

Finally, Fig.~\ref{fig:loop} shows the energy dispersion Eq.~\eqref{eq:Eloop} in the complex plane. The dispersion curve $\varepsilon_{\rm pair}(K)$ as a function of $K$ has a characteristic loop with a  winding number of {unity}, { as can be expected \cite{Zhang_2020} from a single band of edge-concentrated states in Fig.~\ref{fig:1D}}. The extent of the loop is regulated by the complex $K$-dependent  term $U(K)$.
Meanwhile, the energy spectrum for the open boundary conditions, shown in Fig.~\ref{fig:loop} by  colored circles, forms a single curve.  The circle color represents the localization degree of the eigenstates,  quantified by the inverse participation ratio (IPR), defined as $\sum_{n}|\psi_n|^4/[\sum_{n}|\psi_n|^2]^2$. 
More localized eigenstates (brighter circles) correspond to the center of the loop in the periodic structure. Such a loop collapse into a curve is a hallmark of the NHSE. We have also checked that the results do not qualitatively depend on $N$ and $\sigma$ once the effective broadening of the momenta in the finite system $\pi/N\gtrsim\sigma$~\cite{Note1}. Thus, our effective single-composite-particle model links the localization observed for the two-particle bound states in Fig.~\ref{fig:bound-dispersion} to the NHSE.

{\it Summary and outlook.} To summarize, we have demonstrated that the dissociation of bound pairs in a chiral structure can result in the non-Hermitian skin effect, where the pairs are localized at the edge. We illustrated the mechanism for a specific  platform of waveguide quantum electrodynamics, where 
 an array of atoms is coupled to photons propagating in a simple waveguide.
We expect that the effect could be flexibly tuned by controlling the photon dispersion law in structured waveguides, that are experimentally available~\cite{Kim2021}.
The coexistence of the continuum spectrum with the quasiparticle dispersion branch is a generic feature  of various many-body systems, for example, with plasmonic or magnonic excitations~\cite{mahan2013many}.  Once the time-reversal symmetry is broken, the quasiparticle branch acquires a unidirectional part. Therefore, 
we believe that our results could be generalized beyond the waveguide QED setup. {The simplest example could be an extended Bose-Hubbard model, that describes a chain of coupled Kerr-nonlinear cavities with extra nonlocal Kerr nonlinearity~\cite{Gorlach2017b}. The model in Ref.~\cite{Gorlach2017b} is nonchiral, but features bound states hybridized with a continuum. We expect that if extra chiral two-photon tunneling is added, it should also reproduce the considered effect.}

We thank Ekaterina Vlasiuk and Janet Zhong for valuable discussions. 
 The work of A.N.P. has been supported by research grants from the  Minerva Foundation, from the Center for New Scientists, and the Center for Scientific Excellence at the Weizmann Institute of Science.

%

\newpage\clearpage

\onecolumngrid
\setstretch{1.8}
\setlength{\parindent}{1em} 
\setlength{\parskip}{1em}
\large

\setcounter{figure}{0}
\renewcommand{\thefigure}{S\arabic{figure}}
\setcounter{equation}{0}
\renewcommand{\theequation}{S\arabic{equation}}
\let\addcontentsline\oldaddcontentsline

\begin{center}
\bf \large Supplementary Information
\end{center}
\section{Photon dispersion dependence on the chirality}


\begin{figure}[t!]
\centering
 \includegraphics[width=0.99\linewidth]{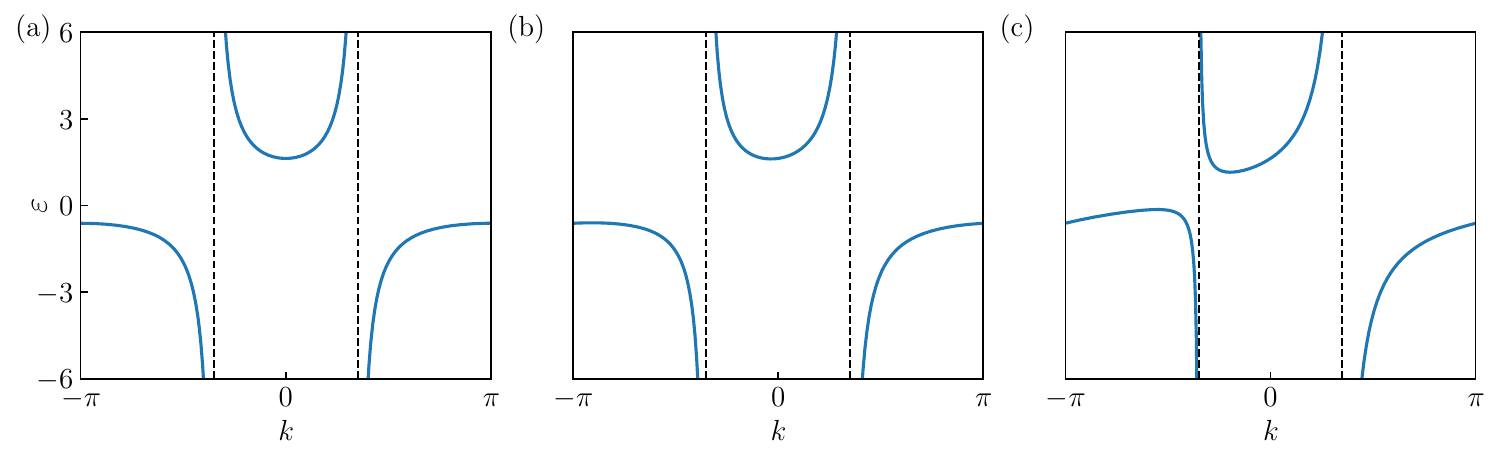}
\caption{ Dispersion of the single-photon excitations.
The calculation has been performed 
 using Eq.~\eqref{eq:single} for (a) $\xi=1$, (b) $\xi=0.7$, 
 (c) $\xi=0.1$, and  $\varphi=0.35\pi$ and 
 $\gamma_{\rm 1D}=1$.  
}
\label{fig:xi-dep1}
\end{figure}

\begin{figure}[b!]
\centering
\includegraphics[width=0.99\linewidth]{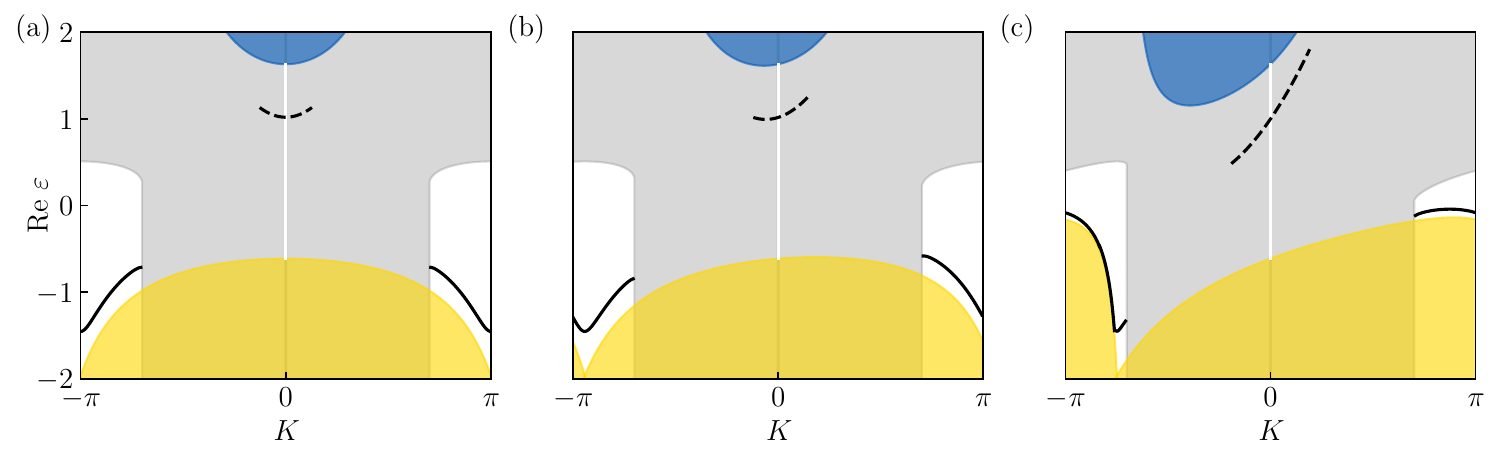}
\caption{ Dispersion of the two-photon excitations. Dashed lines represent the quasi-bound states around $K = 0$, which are extensively discussed in Ref.~\cite{bakkensen2021}.  The calculation has been performed using 
Eq.~(4) in the main text 
for (a) $\xi=1$, (b) $\xi=0.7$, (c) $\xi=0.1$, for $\varphi=0.35\pi$ and 
 $\gamma_{\rm 1D}=1$.  
}
\label{fig:xi-dep}
\end{figure}

Figure~\ref{fig:xi-dep1} shows the single-photon dispersion branch, calculated  for
 three different values of the  chirality parameter $\xi=1,0.7,0.1$ using the following equation
 ~\cite{Sheremet2023}
 \begin{equation}
     \omega(K)=\omega_0+\gamma_{\rm 1D}\frac{\sin\varphi+\theta\sin K}{\cos K-\cos\varphi},\text{ where } 
\theta=\frac{1-\xi}{1+\xi}\:. \label{eq:single}
 \end{equation}
Figure~\ref{fig:xi-dep}(a--c) presents the two-photon dispersion calculated for three different values of  $\xi=1,0.7,0.1$. The continuum of the scattering states is obtained by taking the value range of \[\varepsilon_{\rm scat}(K) = \frac{\omega(q)+\omega(K-q)-2\omega_0}{2}\] for varying $q$. The bound states spectra are obtained by numerically diagonalizing Eq.~(4) in the main text. The middle panel is the same as the inset in Fig.~2(a) in the main text.
The calculation in Figure~\ref{fig:xi-dep} demonstrates, that the binding energy of the considered bound state dispersion branch decreases for smaller $\xi$. The branch approaches the continuum of scattering states, corresponding to two polaritons in the lower dispersion branch (yellow area), see in particular Fig.~\ref{fig:xi-dep}(c). As a result, the bound state stops being localized. This leads to suppression of the considered NHSE, where the bound state is localized at the ``wrong'' left edge of the structure for $\gamma_\to>\gamma_{\leftarrow}$.

The dashed curve in Fig.~\ref{fig:xi-dep} shows the second bound state dispersion branch. This is, in general, a resonant state branch that is hybridized with the continuum and has a finite lifetime. Only in the limit of $\xi\to 0$ it stops being a resonant state and becomes the true bound state~\cite{bakkensen2021,Calajo2022}.

\begin{figure}[tb!]
\centering
\includegraphics[width=0.5\linewidth]{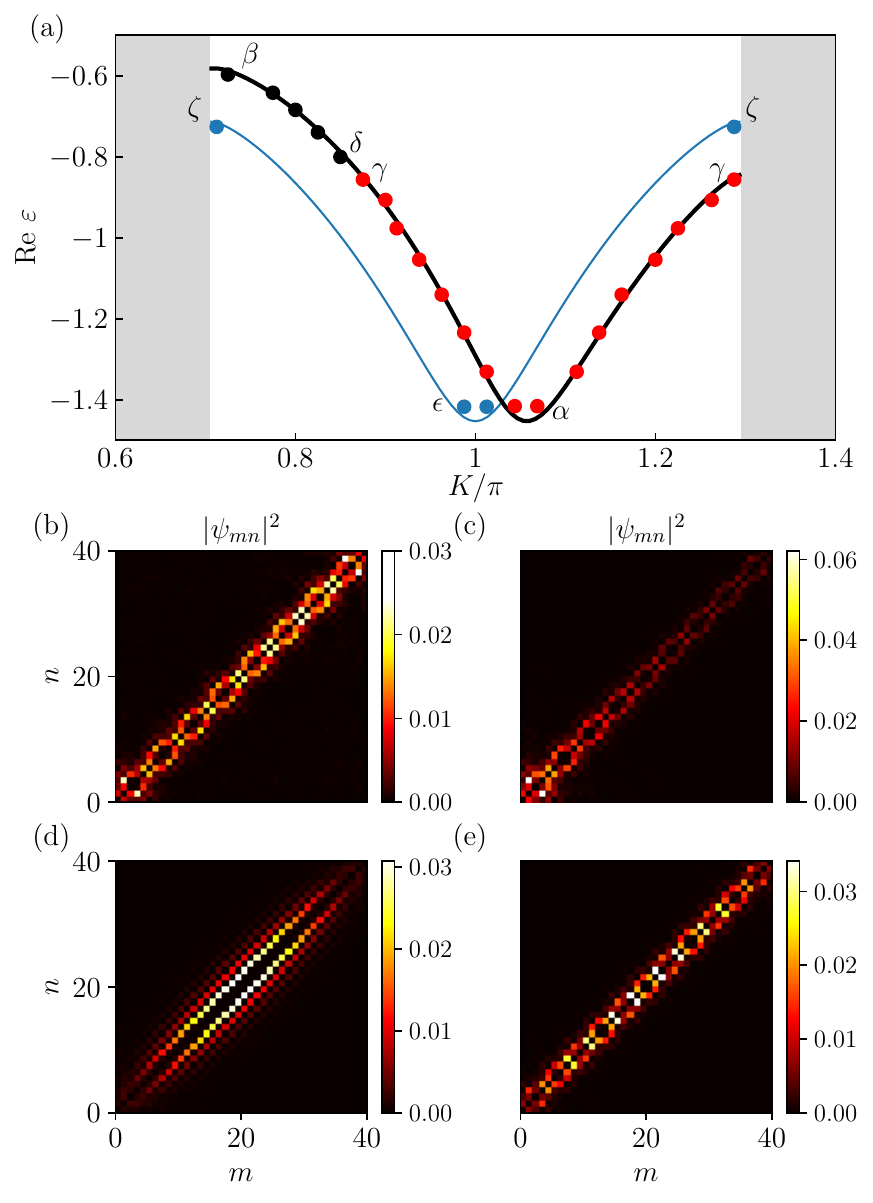}
\caption{Similar to Fig.~2 in the main text, but with more eigenstate examples. (a) Bound pair dispersion law $\varepsilon_{\rm pair}(K)$ in the chiral (thick black curve) and non-chiral (thin blue curve) structure. (b,c) Real-space distributions for states $\gamma$, $\delta$ from the chiral dispersion. (d,e) Real-space distributions for states $\epsilon$, $\zeta$ from the non-chiral dispersion.  The calculation  has been performed for the same parameters as Fig.~2 in the main text: $\varphi=0.35\pi$, $N=40$, $\xi=0.7$ and $\gamma_{\rm 1D}=1$.
}
\label{fig:more_states}
\end{figure}

We also show in Fig.~\ref{fig:more_states} more examples of the eigenstates spatial distribution in chiral (b,c) and nonchiral structures  (d,e). The calculation demonstrates that the bound pair is localized at the edge only if its energy is in the unidirectional range of the chiral structure, Fig.~\ref{fig:more_states}(c).

\section{ Linear-in-K dispersion terms}
\renewcommand{\epsilon}{\varepsilon}
In this section, we {derive an approximate analytical solution} for the linear-in-$K$ terms in the bound state dispersion law
\begin{equation}\label{eq:epair1}
    \varepsilon_{\rm pair}(K)\approx
    \varepsilon_{\rm pair}(\pi)+\alpha(K-\pi)+\frac{(K-\pi)^2}{2m}\:
\end{equation}
near the center-of-mass Brillouin zone edge $K=\pi$.
 
In an infinite periodic array, the wavefunction that describes the double-excited states $\sum_{r,s=1}^\infty \Psi_{rs} \sigma_r^\dag \sigma_s^\dag |0\rangle $ can be written as 
\begin{equation}
    \Psi_{rs} = \e^{\rmi K (r+s)/2} {\chi}_{r-s}, \chi_0 = 0\:.
\end{equation}
The Hamiltonian ~(2) from the main text then assumes the form
\begin{align}
    &\sum_{s=-\infty}^\infty \mathcal{H}_{r,s}(K) \chi_s = 2\varepsilon(K) \chi_r, \chi_0 = 0\:, \\
    &\mathcal{H}_{r,s}(K) = -i\gamma_{\rm 1D} \cos(\frac{K(r-s)}{2}) \e^{\rmi \varphi|r-s|} - \frac{\gamma_{\rm 1D} \epsilon}{2} \sin(\frac{K|r-s|}{2}) \e^{\rmi \varphi |r-s|} \equiv H_0 + H_\epsilon \:,
\end{align}
where $\epsilon = 2\frac{1-\xi}{1+\xi}$ is a parameter characterizing the degree of chirality. We can obtain the analytical expressions of the bound state and scattering states at $K=\pi$ in a non-chiral case~\cite{Poddubny2019quasiflat}:
\begin{equation}
    \chi_{\pm 2r}^{(0)} = \frac{(-1)^r}{\sqrt{2}} \e^{-(r-1)\kappa}\sqrt{1-\e^{-2\kappa}}\:, \quad
    \chi_{\pm(2r+1),q} = \cos q\left(r+\frac{1}{2}\right)\:,
\end{equation}
where $\kappa = -\ln \cos2\varphi$ is the inverse effective size of the bound pair and $-\pi < q \leq \pi$ is the wave vector of relative motion of the two particles in a scattering state. The wavefunction of the bound state is normalized as $\sum_{r=-\infty}^\infty |\chi_{r}^{(0)}|^2=1$.
The bound state and scattering states have the energies
\begin{equation}
    \epsilon_\pi^{(0)} =  2\gamma_{\rm 1D} \cot2\varphi\:, \quad \epsilon(q) = \gamma_{\rm 1D} \frac{\sin\varphi \cos \varphi}{\sin^2\varphi-\cos^2\frac{q}{2}}\:.
\end{equation}

The linear-in-$K$ term, proportional to $\alpha$ in Eq.~\eqref{eq:epair1}, is induced by the chirality of the Hamultonian. It  can be obtained with perturbation theory by considering the first order derivative of the bound state energy at $K=\pi$:
\begin{equation}\label{eq:chiral1}
	\begin{split}
		\frac{d\varepsilon}{dK} &= \langle \chi |\frac{dH}{dK}|\chi\rangle \\
		&= \langle \chi^{(0)} |\frac{dH_\epsilon}{dK}|\chi^{(0)}\rangle +  2\langle \delta\chi |\frac{dH_0}{dK}|\chi^{(0)}\rangle  \\
		&= \langle \chi^{(0)} |\frac{dH_\epsilon}{dK}|\chi^{(0)}\rangle + 2 \int\limits_{-\pi}^{\pi} \frac{dq}{2\pi} \frac{\langle \chi^{(0)}|H_\epsilon^*|\chi_q\rangle }{E_\pi^{(0)}-E(q)} \langle \chi_q|\frac{dH_0}{dK} |\chi^{(0)}\rangle\:.
	\end{split}
\end{equation}
Here, $|\delta \chi\rangle$ is the first order perturbation to  $|\chi^{(0)}\rangle$.
We  remind that the wavefunction amplitudes $\chi^{(0)}$ and $\chi_q$  are real-valued, and that $\langle \chi^{(0)}|\frac{d H_0}{dK}|\chi^{(0)}\rangle = 0$. 

The first term in Eq.~\eqref{eq:chiral1} gives:
\begin{equation}
	\begin{split}
		\langle \chi^{(0)} |\frac{dH_\epsilon}{dK}|\chi^{(0)}\rangle &= -\frac{\gamma_{\rm 1D}\epsilon}{2} \sum_{t,u=-\infty}^{\infty} \chi_s (\frac{|t-u|}{2} \cos(\frac{t-u}{2} \pi) \e^{\rmi \varphi |t-u|}) \chi_t \\
		&= -\frac{\gamma_{\rm 1D}\epsilon}{2} (1-\e^{-2\kappa}) \sum_{r,s=1}^{\infty} [(r+s)\e^{\rmi 2\varphi(r+s) } \e^{-(r+s-2)\kappa} + |r-s|\e^{\rmi 2\varphi|r-s| } \e^{-(r+s-2)\kappa}] \\
		&= \gamma_{\rm 1D}\csc^2 2\varphi \:.
	\end{split}
\end{equation}

The calculation of the second term in Eq.~\eqref{eq:chiral1} is also straightforward but relatively cumbersome. At the first step we obtain
\begin{equation}
	\begin{split}
		\langle \chi_q|\frac{dH_0}{dK} |\chi^{(0)}\rangle &= \rmi \gamma_{\rm 1D}\sqrt{2} \sqrt{1-e^{-2\kappa}} \sum_{r=1}^\infty \sum_{p=0}^\infty \cos[\kappa(p+1/2)] (-1)^r e^{-(r-1)\kappa}\\
		 &\times [\sin\frac{\pi(2r-2p-1)}{2} \frac{2r-2p-1}{2} \e^{\rmi \varphi |2r-2p-1|} + \sin\frac{\pi(2r+2p+1)}{2} \frac{2r+2p+1}{2} \e^{\rmi \varphi(2r+2p+1)}]  \\
		 &= -8\gamma_{\rm 1D} \sqrt{2} \frac{(\cos^2\varphi +\sin^2 q/2)\cos\varphi \cos (q/2) \sin^2\varphi}{(\cos q + \cos2\varphi)(4\cos2\varphi \cos q + \cos 4\varphi + 3)} \:,
	\end{split}
\end{equation}
and
\begin{equation}
	\begin{split}
		2\langle \chi^{(0)}|H_\epsilon^*|\chi_q\rangle &= -\epsilon \gamma_{\rm 1D}\sqrt{2} \sqrt{1-\e^{-2\kappa}}  \sum_{r=1}^\infty \sum_{p=0}^\infty \cos[\kappa(p+1/2)] (-1)^r e^{-(r-1)\kappa} \\
		&\times [ \sin\frac{\pi|2r-2p-1|}{2} \e^{-\rmi\varphi|2r-2p-1|} + \sin\frac{\pi(2r+2p+1)}{2} \e^{-\rmi \varphi(2r+2p+1)}] \\
		&= 4\epsilon\gamma_{\rm 1D} \sqrt{2} |\sin 2\varphi|\frac{ \cos\varphi \bigl[ \left( 2 \cos 2 \varphi + \cos 4 \varphi + 3 \right) \cos (q/2) + 2 \cos 2\varphi \cos (3q/2) \bigr]}{\left( 4 \cos 2\varphi \cos q + \cos 4\varphi + 3 \right)^2} \:.
	\end{split}
\end{equation}
The integral is evaluated as
\begin{equation}\label{key}
	\begin{split}
		&\frac{\epsilon\gamma_{\rm 1D}}{2\pi}\int_{-\pi}^{\pi} dq \frac{8 \sin ^42p \cos \left(\frac{q}{2}\right) (\cos 2p-\cos q+2) \left[2 (2 \cos 2p+\cos 4p+3) \cos \left(\frac{q}{2}\right)+4 \cos 2p \cos (3q/2)\right]}{(4 \cos 2p \cos q+\cos 4p+3)^4}\\
		&=-\epsilon\gamma_{\rm 1D} \sin ^2\varphi (-\cos 2 \varphi+\cos 4 \varphi+4) \csc ^42 \varphi \:.
	\end{split}
\end{equation}
Finally, the result reads
\begin{equation}
    \alpha = \frac{\gamma_{\rm 1D}\cos 3\varphi}{8 \cos^5 \varphi} \epsilon \:,
\end{equation}
this is equivalent to Eq.~(6) in the main text for $\xi\approx 1$.
In a similar way, we can get the effective mass at zeroth order in $\epsilon$~\cite{Poddubny2019quasiflat}:
\begin{equation}
    \frac{1}{m} = - \frac{\gamma_{\rm 1D} \sin\varphi \cos 3\varphi}{8 \cos^6 \varphi} \:.
\end{equation}

\newpage\clearpage
\section{Eigenstates of the effective model}
Here, we analyze in more detail  the effective non-Hermitian Hamiltonian Eq.~(7) in the main text.

\subsection{Nonlocal potential}
\begin{figure}[bt!]
\centering
\includegraphics[width=0.5\linewidth]{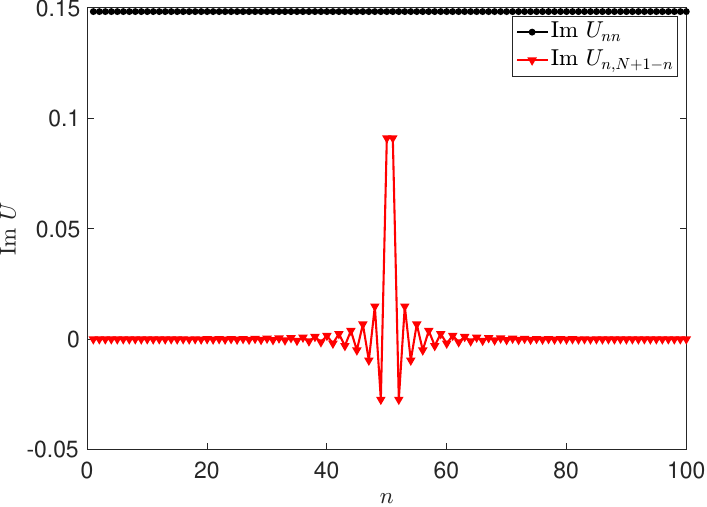}
\caption{Potential $U$ of the effective model.  Black dots and red triangles show the averaged potential $\Im U_{nn}$ and the nonlocal part
$\Im U_{n,N+1-n}$, respectively. Calculation has been performed for $\sigma=0.05$, $\varphi=\pi/2$ and $N=100$, same as Fig.~3 in the main text.
}
\label{fig:U}
\end{figure}
The potential $U_{nm}$ of the effective model, described by Eqs.~(8,9) in the main text, is strongly nonlocal. We illustrate this by plotting in Fig.~\ref{fig:U} the cross-sections of the $\Im U_{nm}$ matrix along the diagonal (black circles) and along the antidiagonal (red triangles). While the diagonal part describes the uniform constant loss, the anti-diagonal part oscillates with the distance between the sites $n$. The oscillation frequency is given by the parameter $\varphi=\pi/2$. It is these oscillations that provide momentum-dependent loss for $|K|<\varphi$.

\subsection{Eigenstates localization}
\begin{figure}[tb!]
\centering
\includegraphics[width=0.65\linewidth]{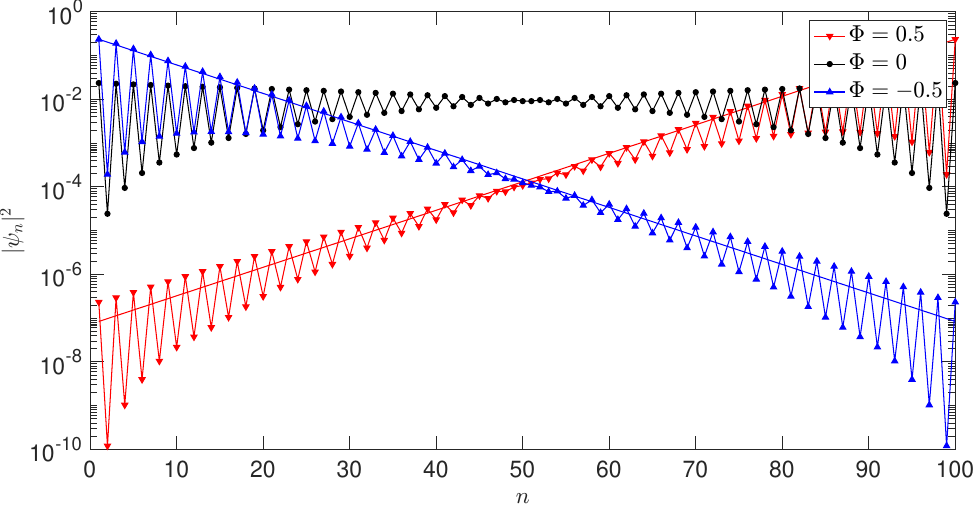}
\caption{
Eigenstate profile of the effective single-particle model for
$\Phi=0$ (circles) and 
$\Phi=\pm 0.5$ (triangles). The thin line shows the analytical dependence
$|\psi_n|^2\propto \exp(\pm \gamma n/2t)$ for $\Phi\ne 0$. 
For each value of $\Phi$ we have selected the  eigenstate with the real part of the energy closest to 0.
Other calculation parameters are: $t=1$,
$2\phi = \pi/2$,
$\gamma = 0.3$, 
$\sigma = 0.05$, and $N=100$.
}
\label{fig:SState}
\end{figure}

\begin{figure}[bt!]
\centering
\includegraphics[width=0.6\linewidth]{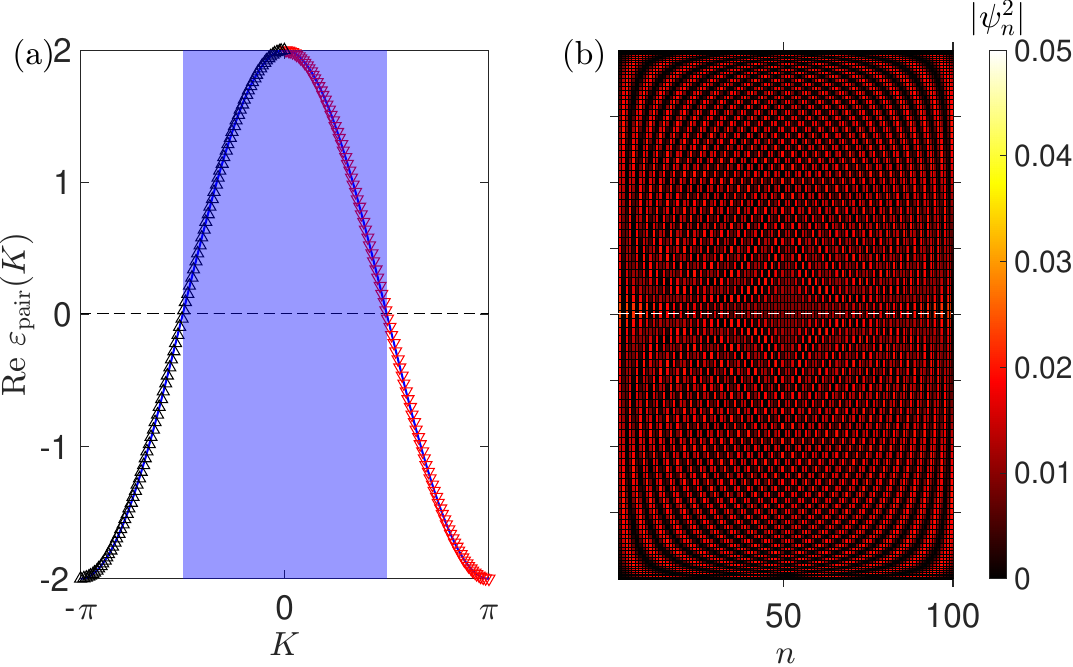}
\caption{Same as Fig.~3 in the main text, but for the non-chiral case, $\Phi=0$.
(a) Solid line shows the dispersion law $\Re\varepsilon_{\rm pair}(K)$ calculated  following Eq.~\eqref{eq:Eloop1}.  The shaded area illustrates the range of momenta $-\varphi<K<\varphi$ with a large loss.
Triangles show the momenta corresponding to the eigenstates in the finite structure with the open boundary conditions.
(b) Spatial probability distribution for all the eigenstates in the finite structure.
Calculation has been performed for the following values of parameters: $t=1$,
$\Phi = -0.5$, 
$2\phi = \pi/2$,
$\gamma = 0.3$, 
$\sigma = 0.05$, and $N=100$.
}
\label{fig:S1DNonChiral}
\end{figure}
To verify the exponential character of localization of the eigenstates of the effective model, we have plotted in Fig.~\ref{fig:SState} the most-localized state, corresponding to the energy values in the center of the band, that is, $\Re \varepsilon_{\rm pair}$ closest to zero [see Fig.~4 in the main text]. The calculation demonstrates that the wavefunction is not localized at the edge for $\Phi=0$ (red circles). For $\Phi\ne 0$ it is exponentially at the left or right edge depending on the sign of $\Phi$ (triangles). The localization length can be estimated by solving the dispersion equation 
\begin{equation}\label{eq:Eloop1}
    \varepsilon_{\rm pair}(K)=2t\cos(K+\Phi)-\rmi U(K).
\end{equation}
for $\varepsilon_{\rm pair}=-\rmi\gamma/2$ and $U=\gamma$ (as can be inferred from Fig.~4 in the main text
). The result is 
$|\Im K|=\mathop{\mathrm{arcsinh}}(\gamma/4t)\approx \gamma/(4t)$ for $\gamma\ll t$. Thin straight lines in Fig.~\ref{fig:SState} show the corresponding exponential dependences  and well describe the overall decay of the numerically calculated eigenmode.

Figure~\ref{fig:S1DNonChiral} shows the dispersion (a) and  the spatial profile (b) for the eigenstates in the effective single-particle model Eq.~(7) in the main text. The parameters correspond to Fig.~3 in the main text, but with zero chirality, $\Phi=0$. As a result, the eigenstates in Fig.~\ref{fig:S1DNonChiral}(b) do not manifest localization at the edge.

\begin{figure}[t!]
\centering
\includegraphics[width=0.55\linewidth]{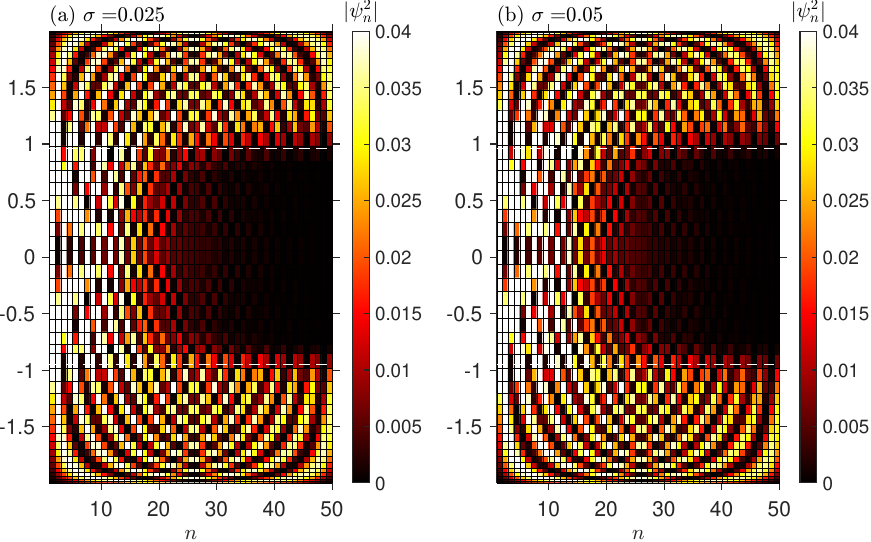}
\caption{Same as Fig.~3b in the main text, but
for $N=50$ and  $\sigma=0.025$ (a) and $\sigma=0.05$ (b).
} 
\label{fig:StatesProfileSigma}
\end{figure}

\begin{figure}[bt!]
\centering
\includegraphics[width=0.53\linewidth]{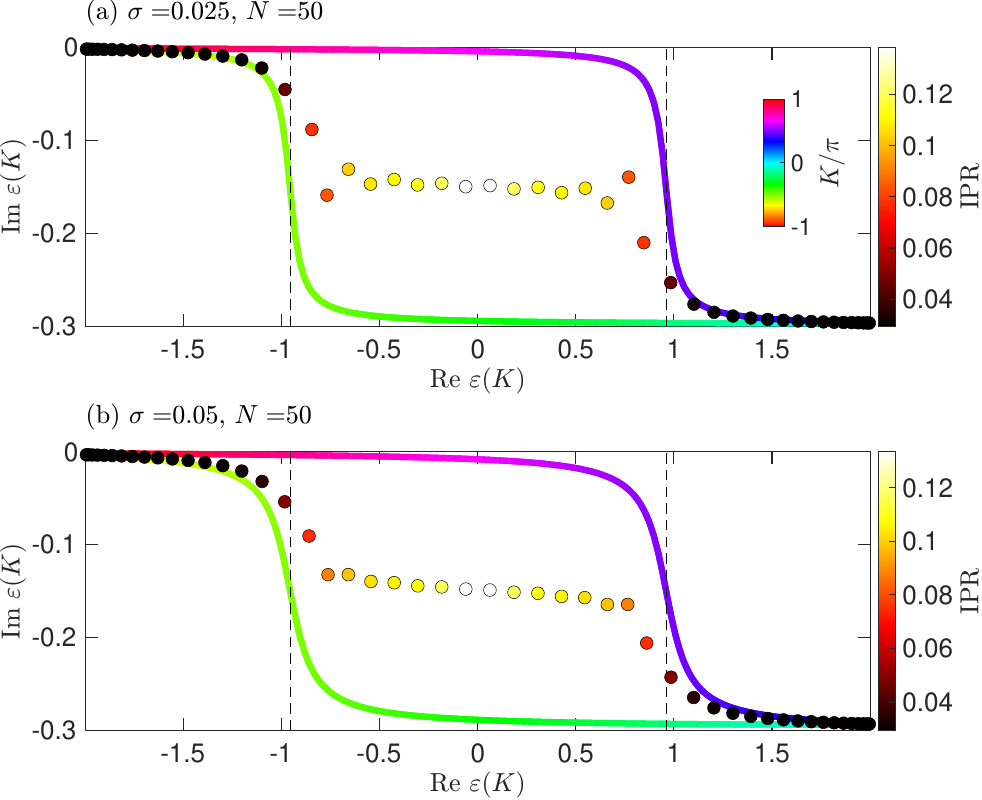}
\caption{Same as Fig.~4 in the main text, but
for $N=50$ and  $\sigma=0.025$ (a) and $\sigma=0.05$ (b).
Colored curve shows the complex pair energy dispersion $\varepsilon_{\rm pair}(K)$ calculated following Eq.~\eqref{eq:Eloop1}  Filled circles present the energies of the eigenstates for the finite structure. The circle's color encodes the eigenstate's inverse participation ratio (IPR). 
} 
\label{fig:SSigma}
\end{figure}

To check that the results do not qualitatively depend on  $N$ and $\sigma$ for $\pi/N\gtrsim \sigma$ we present in Fig.~\ref{fig:StatesProfileSigma} the same eigenstates spatial profiles as in Fig.~3b in the main text, but calculated for $N=50$ and two values of the step smoothness parameter $\sigma=0.025$ (a) and $\sigma=0.05$ (b).  Figure~\ref{fig:SSigma} shows the complex energy dispersion, just as in Fig.~4 in the main text.


\begin{figure}[b!]
\centering
\includegraphics[width=0.47\linewidth]{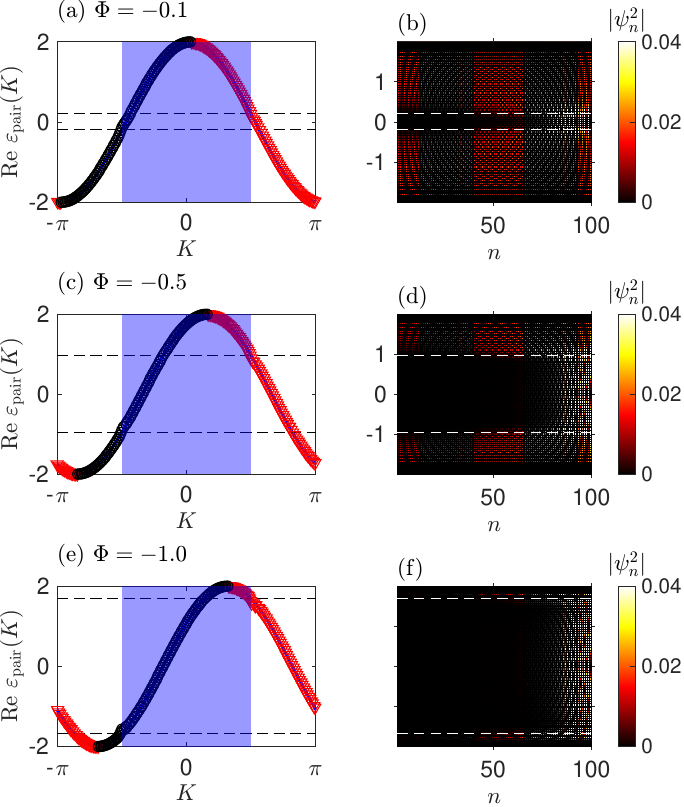}
\caption{Same as Fig.~4 in the main text, but
for three different values of the chirality parameter $\Phi=-0.1$ (a,b),
$\Phi=-0.5$ (c,d) and $\Phi=-1$ (e,f).
(a) Solid line shows the dispersion law $\Re\varepsilon_{\rm pair}(K)$ calculated  following Eq.~\eqref{eq:Eloop1}.  Blue shaded area illustrates the range of momenta $-2\varphi<K<2\varphi$ with a large loss.
Triangles and circles show the momenta corresponding to the eigenstates in the finite structure with the open boundary conditions.
(b) Spatial probability distribution for all the eigenstates in the finite structure.
Horizontal dashed lines in (a,b) indicate the unidirectional range of energies corresponding to the solutions concentrated at the edge.
}
\label{fig:ScanAlpha}
\end{figure}

Figure~\ref{fig:ScanAlpha} examines the dependence of the spectrum and the eigenstates on the chirality parameter $\Phi$. As expected, the states become stronger localized at the edge for increasing $|\Phi|$. The width of the unidirectional spectral range increases as well. For $\Phi=-1$, this range occupies almost the whole band, see Fig.~\ref{fig:ScanAlpha}(e,f).

\end{document}